\documentclass[smallcondensed]{svjour3}
\smartqed
\usepackage{cite}
\usepackage{amsmath,amssymb,amsfonts,dsfont}
\usepackage{algorithmic}
\usepackage{times}
\usepackage{moreverb}
\usepackage{mathrsfs}
\usepackage{bm}
\usepackage{url}
\usepackage{makecell}
\usepackage[T1]{fontenc}
\usepackage{nicematrix}
\usepackage{graphicx}
\usepackage{hyperref}
\usepackage{textcomp}
\usepackage{xcolor}
\usepackage{pifont}
\usepackage{textpos}
\usepackage{multirow}

\def\BibTeX{{\rm B\kern-.05em{\sc i\kern-.025em b}\kern-.08em
    T\kern-.1667em\lower.7ex\hbox{E}\kern-.125emX}}

\newcommand{\chen}[1]{\textcolor{black}{#1}}
    
\begin{document}

\title{Trustworthy Federated Learning: Privacy, Security, and Beyond}

\author{Chunlu Chen\footnotemark[6]\footnotemark[4],
        Ji Liu\footnotemark[1]\footnotemark[5]\footnotemark[4],
        Haowen Tan\footnotemark[9],
        Xingjian Li\footnotemark[11],
        Kevin I-Kai Wang\footnotemark[7],
        Peng Li\footnotemark[8],
        Kouichi Sakurai\footnotemark[6],
        Dejing Dou\footnotemark[10]
}

\date{Received: date / Accepted: date}

\institute{
\footnotemark[1] Corresponding author. \\
\footnotemark[4] Equal contribution. \\
\footnotemark[5] HiThink Research, Hangzhou, China.\\
\footnotemark[11] Carnegie Mellon University,  Pittsburgh, United States.\\
\footnotemark[6] Information Science and Electrical Engineering Department, Kyushu University, Fukuoka, Japan.\\
\footnotemark[7] Electrical, Computer and Software Engineering Department, The University of Auckland, Auckland, New Zealand.\\
\footnotemark[8] Computer Science and Engineering Department, The University of Aizu, Aizuwakamatsu, Japan. \\
\footnotemark[9] Information Science and Technology Department, Zhejiang Sci-Tech University, Hangzhou, China. \\
\footnotemark[10] Fudan University, Shanghai and BEDI Cloud, Beijing, China. 
}

\authorrunning{Chen and Liu et al.}

\maketitle

\begin{textblock*}{8cm}(3cm,15cm) % {block width} (coords) 
   {\huge To appear in KAIS}
\end{textblock*}

\begin{abstract}
\chen{While recent years have witnessed the advancement in big data and Artificial Intelligence (AI), it is of much importance to safeguard data privacy and security. As an innovative approach, Federated Learning (FL) addresses these concerns by facilitating collaborative model training across distributed data sources without transferring raw data. However, the challenges of robust security and privacy across decentralized networks catch significant attention in dealing with the distributed data in FL. In this paper, we conduct an extensive survey of the security and privacy issues prevalent in FL, underscoring the vulnerability of communication links and the potential for cyber threats. We delve into various defensive strategies to mitigate these risks, explore the applications of FL across different sectors, and propose research directions. We identify the intricate security challenges that arise within the FL frameworks, aiming to contribute to the development of secure and efficient FL systems.}

\end{abstract}

\section{Introduction}

In recent years, rapid advancements in big data and Artificial Intelligence (AI) technologies have ushered in an era characterized by an unprecedented proliferation of interconnected Internet of Things (IoT) devices and web platforms. This digital tapestry, while instrumental in catalyzing the data revolution, concurrently yields vast quantities of distributed data — a significant portion of which is sensitive in nature. Notably, there exists a gap in the adequate protection of this sensitive information, a critical oversight in the current data-centric world.
The emergent challenges have not gone unnoticed at the legislative level. A myriad of regulations, including such as the Cybersecurity Law of the People’s Republic (CLPR) of China \cite{CCL}, the General Data Protection Regulation (GDPR) \cite{GDPR}, the California Consumer Privacy Act (CCPA) \cite{CCPA}, and the Consumer Privacy Bill of Rights (CPBR) \cite{Gaff2014}, have been established to safeguard the privacy and security of raw data. Current estimations indicate that these privacy legislations may encompass up to 75\% of the global population \cite{Gartner}, necessitating over 80\% of worldwide enterprises to conform by the culmination in 2023.
In this dynamic landscape, the development and deployment of sophisticated defense methodologies are imperative. Such techniques are pivotal to maintain data privacy and security throughout the lifecycle of machine learning models, including both training and inference phases.

Traditional centralized machine learning paradigms necessitate the aggregation of data at a single server or data center, serving as the nexus for both training and inference operations. Training, an iterative process, refines machine learning model parameters through specific algorithms and can be computationally intensive and time consuming \cite{wang2019distributed}. In contrast, the inference phase leverages these trained models to deduce predictions or classifications \cite{2021FromDM}.
The introduction of Distributed Machine Learning (DML) techniques augments both the accuracy and computational efficiency of the model training process. However, this decentralization inevitably  exacerbates concerns regarding data privacy and security. Federated Learning (FL) emerges as a pivotal solution in this context. Instead of transferring raw data, which incurs potential privacy violation, FL facilitates the dissemination of a global model to individual devices. These devices, in turn, harness local data to refine the model. Post-training, the local devices relay the updated model parameters to the central server for amalgamation. This iterative process is perpetuated until model convergence is achieved, ensuring data remains local, thereby supporting privacy and security.
While FL offers clear advantages, it also faces challenges. Weak points in the communication links between devices and central servers can lead to cyberattacks. In addition, if servers or devices are compromised, they might bring in malicious activities, threatening the overall security of the system.

FL has emerged as a key development in the field of modern DML. Numerous surveys delve into the fundamental aspects of FL, discussing topics like deployment architecture, system lifecycle, defining characteristics, classifications, and the range of open-source tools available \cite{2021FLSurvey, QinbinLi2021ASO, 2020FederatedLA, 2021FromDM, wen2023survey}. Recent studies analyze FL within the software engineering domain, providing insights into the detailed processes involved in developing FL systems \cite{2021Software, supriya2023survey}.
In terms of application, significant works focus on specific application of FL. Important studies in this realm cover multiple topics, such as edge computing \cite{QiXia2021ASO}, integration methods for the IoT and the Industrial IoT (IIoT) \cite{2021IoTSurvey, MParimala2021FusionOF, VirajKulkarni2020}, strategies centered on personalization \cite{2020TowardsUU}, and in-depth reviews exploring the economic impact of FL adoption \cite{Zhou2021ASO}.
In terms of security, a plethora of seminal works have structured frameworks that elucidate the intricacies of FL security \cite{MOTHUKURI2021619, zhang2023survey}. Alongside these, there are focused analyses identifying potential security risks, with an emphasis on the security and privacy challenges within FL \cite{2020ThreatsTF, ratnayake2023review}. The growing concerns about privacy breaches are a significant topic of interest in recent discussions \cite{XuefeiYin2021ACS, rodriguez2023survey}. A consistent finding across these studies is the presence of challenges during the development and deployment of FL, especially concerning system vulnerabilities and device reliability \cite{2021Kairouz}.

These studies reveal a research shortfall, emphasizing the need for in-depth investigations into security, privacy, and relevant defensive strategies within the FL paradigm.
\chen{Thus, based on the FL system architecture, we explore the security and privacy issues faced at each architecture layer and comprehensively discuss the existing defense techniques designed to enhance the ability to resist various types of security vulnerabilities. The notable contributions of this paper are as follows:}

\begin{itemize}
\item
\chen{We propose a universal FL system architecture that encompasses infrastructure, algorithms, and user services. This architecture aids in the evaluation of existing FL systems. The current literature lacks this holistic view, which distinguishes our work from existing surveys. }
\item
\chen{We provides a comprehensive overview of the security and privacy issues present in FL, as well as the primary attack methods. And also discuss a range of defense techniques against these attacks, offering practical guidance for system developers. }
\item
\chen{We analyze the applications of FL systems and identify future research directions. This contribution enriches the discourse on FL and highlights opportunities for further development.}
\end{itemize}

The remainder of this paper is organized as follows.
In Section \ref{sec:FLOverview}, we introduce the fundamental concepts of FL and elucidate the proposed FL system architecture.
Section \ref{sec:FLSecurity} delves into the prevalent security issues associated with FL. While some threats in FL might arise non-maliciously due to device malfunctions or unpredictable participant behaviors, there are malicious threats that intentionally aim to undermine the system. These can manifest as data poisoning, model corruption, and inference attacks. The decentralized framework of FL offers enhanced privacy in machine learning but also introduces numerous security challenges \cite{gabrielli2023survey}. To address these challenges, we discuss a variety of defense measures. These strategies, crafted considering vulnerabilities from both client devices and central servers, are mainly classified into proactive and reactive types. Proactive defenses aim to preemptively identify and mitigate threats, while reactive defenses come into play once an attack has been detected. Technologies that have been extensively researched in this context include encryption, Differential Privacy (DP), and anomaly detection.
Section \ref{sec:FLApplication} explores the various applications of FL. This section is followed by Section \ref{sec:challenges}, which elucidates challenges and potential future research directions.
Finally, Section \ref{sec:conclusion} provides a conclusion to the paper.

\section{Overview of Federated Learning}
\label{sec:FLOverview}

In this section, we introduce the basic concepts of DML and FL. Then, we discuss the differences between FL and DML. Afterward, we explain the general FL framework architecture.

\subsection{Federated Learning}

Machine learning systems are traditionally reliant on aggregating raw data to a centralized server for model training. A large dataset generally corresponds to high accuracy of a trained model. However, computational bottlenecks emerge when handling extensive data. DML offers a countermeasure to this issue by implementing parallelization and concurrent execution across multiple processing units, including CPUs, GPUs, and TPUs. This not only improves efficiency but also enhances scalability. DML systems can be structured either in a distributed or decentralized fashion. The former utilizes central servers for data interaction and device synchronization, while the latter emphasizes peer-to-peer interactions and node equality. 

As data volumes swell in ML tasks, various concerns, including legal, security, and privacy issues, have emerged. In DML systems, both data and models are potential targets for attacks, with potential outcomes being data or model leaks, or compromise of the availability of trained models. This amplifies the necessity to address these challenges, while FL offers a promising solution.

Emerging as an innovative paradigm, FL mitigates the security and privacy issues inherent in DML \cite{2017communication}. These include concerns related to data leakage during communication \cite{2020Chiu} and the presence of data silos in various industries, an outcome of the challenges associated with data sharing \cite{zhang2021survey,liu2015survey,liu2016multi,liu2018efficient,de2019data}. FL, as a DML methodology, creates global models utilizing virtually aggregated data, without requiring raw data exchange between individual data sources. Instead, it relies on sharing model parameters or intermediate results, such as gradients, between these sources. This attribute allows FL to achieve comparable accuracy to traditional ML methods, while enhancing the security and privacy of raw data.

\chen{Depending on the network topology, FL can be either centralized or decentralized \cite{MOTHUKURI2021619, 2019Qinbin}. In a centralized FL architecture, a central server acts as the hub where the global model is built, managed, and updated. This server collects and aggregates parameters from connected devices, which then use the updated global model \cite{Li2020FedProx,Zhang2022FedDUAP,liu2022Efficient,liu2022MultiJob,JinAccelerated2022,liu2022multi,liu2024efficient}. Although this architecture simplifies management and synchronization, it potentially limits system scalability due to its reliance on a single server, and can impose a heavy communication load in large-scale applications, making it vulnerable to single points of failure and potential bottlenecks \cite{lu2021optimal}.}

\chen{In contrast, a decentralized FL architecture operates without a central server, reducing single point of failure risks and distributing the processing load. Devices or nodes in this setup directly communicate with each other, using a consensus algorithm to ensure trust and reliability \cite{2021FLSurvey, gabrielli2023survey}. Each participant in this network refines their model by sharing information directly with their neighbors, enhancing privacy and reducing the risk of data leakage \cite{Li2022FedHiSyn}. However, this approach significantly increases communication costs as each device must handle multiple connections, and establishing effective consensus among a large number of clients becomes challenging \cite{2021Kairouz}. To further enhance the security and traceability of decentralized FL, recent works have integrated blockchain technology into the architecture \cite{2021Blockchain, 2021Biscotti, 2019FLChain, 2021Secure, qammar2023securing,liu2024enhancing}. This integration leverages blockchain's inherent characteristics of decentralization, transparency, and immutability, providing a robust mechanism for secure and traceable training processes without the need for third-party regulation \cite{li2022blockchain, 2019Poster}. The blockchain-based FL systems can effectively mitigate risks such as data tampering and model poisoning, ensuring a higher level of security in decentralized environments.}

\chen{These different architectures of FL each have distinct security implications. Centralized FL is particularly susceptible to data privacy breaches and centralized control issues, whereas decentralized FL faces challenges related to increased communication overhead and the complexities of maintaining consensus without compromising on the speed or efficiency of the learning process. }

Depending on data distribution, FL can be classified as Horizontally FL (HFL), Vertically FL (VFL) \cite{liu2023distributed}, and Federated Transfer Learning (FTL) \cite{2020IEEEGuide}. In the context of cross-device implementation, with its inherent challenges in maintenance and flexibility, this paper primarily focuses on HFL.

Despite the advantages offered by FL, it is still susceptible to both non-malicious failures and malicious attacks \cite{2021Kairouz}. In FL systems, data and models can become targets for attackers. \chen{An attacker can assume various roles, from a passive spectator to an active participant in the system, each associated with different degrees of potential threat to privacy and security (Table \ref{tab:CompFL}).} An in-depth discussion of privacy and security issues in FL systems is presented in Section \ref{sec:FLSecurity}, detailing how these concerns can be mitigated to ensure the successful application of FL in various domains.

% Table here
\begin{table*}[t]
  \caption{Attacker role and their targets in FL system}
  \label{tab:CompFL}
  \centering
  \scalebox{0.7}{
    \renewcommand{\arraystretch}{1.25}
	\begin{tabular}{|c|c|m{10cm}|}
		\hline
		Attacker Role & Targets & Description \\
    \hline\hline
    Spectator \cite{2020From} & Model or Data & Attackers can collect sensitive information from interactions within systems. \\
    \hline
    \multirow{2}{*}{{Devices \cite{2020Chuan}}} & Model & Attackers may leak or change individual submodel information. \\
\cline{2-3}   & Data & Attackers may leak local training dataset. \\
    \hline
    Servers \cite{2021Kairouz} & Model & Attackers may leak model information or destroy model availability, either global or local models. \\
    \hline
	\end{tabular}
	}
\normalsize
\end{table*}

\subsection{From DML to FL}
\label{sec:Compare}

FL is an offshoot of DML that facilitates collaborative learning without necessitating raw data sharing, thus making significant strides in security and efficiency \cite{2020From, 2021FromDM}. This development is particularly pertinent with the burgeoning trend of IoT and AI technologies. In this section, we investigate the distinct features of DML and FL across four key dimensions: training data, system workflow, security, and fault tolerance.

\begin{itemize}
\item \textbf{Training Data}

In FL, due to the heterogeneity of participating devices \cite{che2022federated,jia2024efficient,liu2024aedfl,liu2024fedasmu}, data distribution and volume can demonstrate remarkable variances among devices. Such diversity leads to a non-Independent and Identically Distributed (Non-IID) distribution in the training data. Additionally, the volume of data on each individual device tends to be asymmetrical, which lead to greater diversity \cite{2018Communication}. Notably, the processing of data occurs directly on these devices, circumventing the requirement for a distinct data management server. 
In contrast, DML assumes that the training data distribution of each worker is a random sample procured from the entire dataset \cite{wang2017machine}. This necessitates a data management server for data collection, feature engineering, and dataset partitioning tasks.

\item \textbf{System Workflow}

The conventional DML workflow comprises four main stages: initiation, training, evaluation, and deployment \cite{wang2017machine}. The initiation stage involves data preprocessing and model initialization, tailored to specific problems, with techniques like Logistic Regression (LR), Support Vector Machine (SVM), and Neural Networks (NN) being predominantly used for classification tasks. The training phase utilizes the curated data to train the model for the specific task. Thereafter, the model undergoes evaluation using test data to gauge its performance and to ascertain its effectiveness. Once the model is approved in this evaluation phase, it gets deployed into the production environment.
FL workflow mirrors that of DML to a large extent, but deviates notably in the initiation phase. In particular, FL forgoes data preprocessing at a central location \cite{zhang2021survey}. The distributed nature of training data across client devices allows each client to undertake independent data preprocessing based on the demands of tasks. The central server has no access to this data, ensuring that data owners retain full control over their devices and data. In contrast, DML provides the central server with comprehensive control over both the training data and the workers.

\item \textbf{Security}

DML poses a higher risk of data leakage as data and model parameters are disseminated to workers via a communication network. Although encryption techniques can attenuate this problem, they also substantially inflate the computational and communication costs of the system\cite{2017Lian}.
FL, in contrast, naturally mitigates data leakage risk as it does not involve sharing raw data and all training occurs locally. However, even in FL, the gradients shared with servers for global model training can potentially disclose information about the training data \cite{2021Kairouz, 2019Deep}. Consequently, several security techniques like encryption and obfuscation should be deployed to bolster the security of system.

\item \textbf{Fault Tolerance}

DML incorporates fault tolerance mechanisms to handle potential issues like worker unavailability due to downtime \cite{2022misbehaviour}. To preempt system delays or failures, it might be necessary to reserve additional resources in advance.
Meanwhile, the fault tolerance mechanism of FL focuses on two distinctive challenges. First, there is the prospect of devices spontaneously dropping out of the system \cite{VirginiaSmith2017}. The second concern relates to Byzantine attacks, where malicious participants intentionally use inaccurate data to train their local models, and then submit these faulty parameters to the server, severely undermining the performance of the global model \cite{2018DRACO, 2021So}. To circumvent these challenges, FL systems may deploy strategies such as allowing for a limited number of offline clients or employing client anomaly detection algorithms to discern potential malicious activities.

\end{itemize}

FL, as a specialized manifestation of DML, facilitates model training without necessitating data exchange, leading to significant enhancements in security and communication efficiency. However, it also presents unique challenges. For instance, malicious clients may attempt to exploit the global model or execute a range of malevolent attacks, such as inference attacks or poisoning attacks. To counteract these threats, integrating anomaly detection mechanisms to identify abnormal clients or updates might be a feasible solution. Yet, one must be aware that the incorporation of multiple defensive mechanisms could inflate communication and computation costs, potentially compromising the overall efficiency of system.

\subsection{System Architecture of FL}

\begin{figure}[t]
    \centering
    \includegraphics[width=0.65\textwidth]{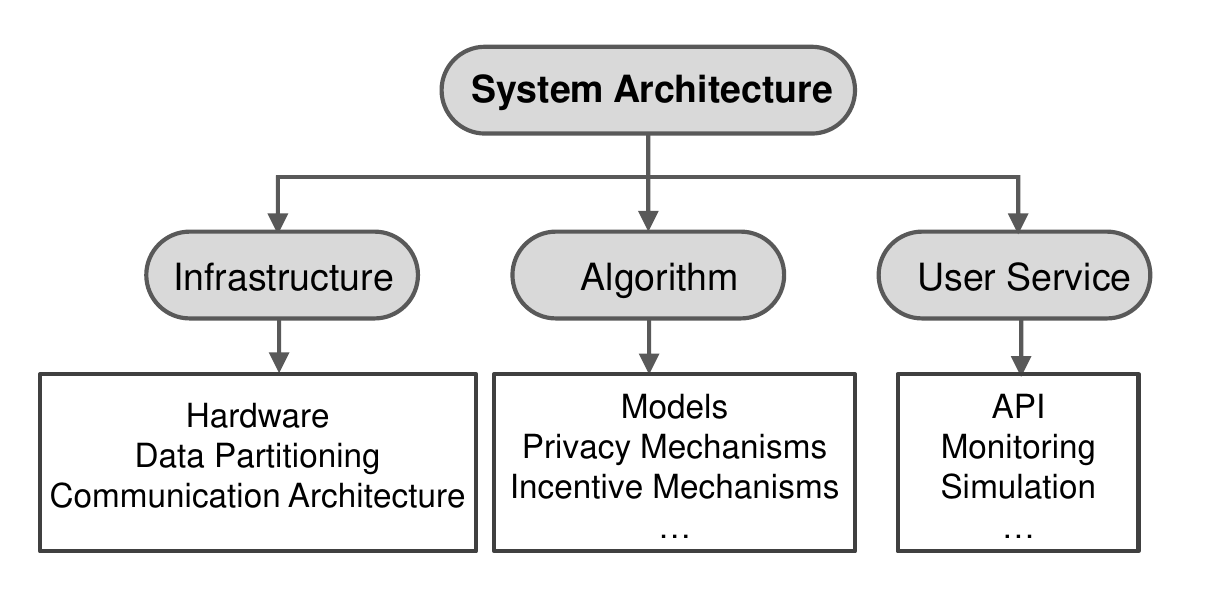}
    \caption{The common FL system architecture}
    %\vspace{-9mm}
    \label{fig:system}
\end{figure}

This section aims to present a comprehensive examination of prevalent open source FL system architectures, encompassing TensorFlow Federated (TFF) \cite{TFF}, PySyft \cite{Pysyft}, FedML \cite{FedML}, Federated AI Technology Enabler (FATE) \cite{FATE}, PaddleFL \cite{PaddleFL}, and Rosetta \cite{Rosetta}. In light of this review, we propose a generalized architecture for FL systems, as depicted in Figure \ref{fig:system}, which comprises three core constituents: the infrastructure, algorithm, and user service.

\begin{itemize}
    \item Infrastructure
    
    The infrastructure component within an FL system encompasses resource management, training data, and the communication architecture. Managing resources to accommodate model training while attenuating computing bottlenecks, such as Graphic Processing Units (GPUs) \cite{2012cuda} and Tensor Processing Units (TPUs) \cite{2017tpu}, is a pivotal challenge in the FL systems. Notably, TFF, PySyft, and FedML offer support for both Central Processing Units (CPUs) and GPUs. Training data in FL is provided by the devices, necessitating a large volume to enhance prediction accuracy, yet the non-uniform and Non-IID nature of training data across devices imposes additional complexity in model training \cite{2018noniid, 2021CFL}. The communication architecture in an FL environment can follow either a centralized or decentralized framework. A centralized architecture involves a central server that aggregates parameters from devices and broadcasts the updated model, while a decentralized architecture permits devices to directly update their models by communicating with neighboring devices.

    \item Algorithm
    
    The algorithm component within an FL system constitutes models, privacy-preserving mechanisms, and incentive mechanisms. Devices in FL systems typically collaborate to train a model that addresses specific machine learning challenges. Neural Networks (NN), tree models, and linear models are widely utilized, each with distinct advantages. NN models have been recognized for their top-tier performance across various applications, such as image classification. However, tree and linear models are often preferred for their comprehensibility and effectiveness. It is noteworthy that NN models typically require labeled data for most tasks. In addition, ensemble methods \cite{shi2021fed} have been proposed to bolster system performance and accuracy by amalgamating multiple models.
    
    Although FL provides a mechanism for training models without raw data sharing, the interaction processes within the system can potentially leak sensitive data or models. To counteract these privacy issues, encryption and obfuscation techniques are extensively implemented in FL systems. Encryption techniques, such as Homomorphic Encryption (HE) and Secure Multi-Party Computation (MPC), protect data during the communication process, whereas obfuscation techniques, including DP \cite{asoodeh2021three,ha2019differential,boulemtafes2020}, introduce noise to safeguard data. Notably, TFF and FedML support DP, FATE supports cryptographic methods, and both PySyft and PaddleFL cater to DP and cryptography.
    
    The role of incentive mechanisms in FL systems is to reward participants for their contributions, encouraging sustained participation and model sharing. These mechanisms employ either positive or negative incentives, which respectively aim to motivate participants through rewards or deter harmful actions by imposing penalties. A fair evaluation metric to gauge the contribution of each participant can attract more participants, and rewards can then be allocated accordingly \cite{2020sustainable, 2020Profit}. 
    Blockchain-based incentive systems have been gaining substantial attention, as they can document participant training activities and compensate active contributors with cryptocurrency \cite{2019FLChain, 2019deepchain, qammar2023securing}. 
    \chen{In addition, to enhance the scalability of the system and reduce performance bottlenecks, we employ blockchain sharding technology to enable parallel model training across multiple shards \cite{Sharding2021Hajar}. This technique divides the blockchain network into several smaller, independent segments (known as "shards"). Each shard independently processes transactions and verifies blocks, thereby increasing the overall network throughput and processing capacity.}
    
    \item  User Service
    
    The user service component within an FL system provides functionalities such as algorithm Application Programming Interfaces (APIs), monitoring tools, and simulation capabilities. Simulations enable rapid efficacy evaluations of an algorithm by mimicking collaborative training across multiple devices. These pre-deployment evaluations expedite the decision-making process for real-world applications. Operational issues can be identified and resolved in a timely manner through monitoring and statistical analysis of the model training process. APIs serve to facilitate algorithm execution and customization to meet user-specific requirements. Most open-source FL systems currently support macOS and Linux platforms, aligning with developer needs. While TFF and PySyft provide comprehensive building blocks for FL process implementation, FATE, Rosetta, PaddleFL, and FedML offer algorithm-level APIs for direct utilization. Moreover, the FATE-Board visualization module provides a graphical depiction of data derived from tracking, statistics, and monitoring of the model training process, enhancing interpretability.

\end{itemize}

\section{Security and Privacy of FL}
\label{sec:FLSecurity}

In the realm of DML, robustness denotes the capacity of system to efficiently counteract or mitigate security threats. FL, characterized by its reliance on distributed devices, enables cooperative model training without necessitating the sharing of raw data. The distributed architecture intrinsic to FL intensifies the intricacy of detecting vulnerabilities and potential adversarial interventions \cite{2021Kairouz}.
The threats to the integrity of an FL system can be principally classified into two categories, i.e., non-malicious failures and malicious attacks. 
Non-malicious failures are inadvertent system impairments emanating from inherent vulnerabilities. Examples encompass device malfunctions, the introduction of overly noisy training datasets, and unpredictable participant behaviors.
Malicious attacks are deliberate infringements orchestrated by adversaries aiming to compromise the FL system. Such attacks encompass data and model poisoning, adversarial manipulations, and inference-based assaults.
While tailored strategies can be designed in advance to counteract malicious incursions, non-malicious setbacks often manifest unpredictably. Remediation for the latter is frequently reactive, necessitating post-incident solution formulations. To bolster the robustness of FL frameworks, it is imperative to meticulously assess the spectrum of potential security breaches and to proactively design and integrate mechanisms that can deter or neutralize them. Subsequently, we delve into a comprehensive discourse on the multifaceted security and privacy concerns inherent to FL architectures.

\subsection{Non-Malicious Failures on FL}

FL systems, despite their distributed promise, grapple with vulnerabilities stemming from both hardware and systemic discrepancies \cite{2020SecureAF}. At the hardware echelon, failures often manifest as a consequence of infrastructural malfunctions. Such malfunctions can be attributed to a myriad of sources, from subpar equipment quality to exogenous factors including environmental calamities or inadequately maintained apparatus.

These inherent system susceptibilities serve as potential gateways for adversaries, enabling them to leverage exploits ranging from malicious code injections to Distributed Denial of Service (DoS) offensives \cite{Berman2019ASO}. Compounding the intricacies are the challenges of noisy training data and participant unreliability \cite{2021Kairouz}. The former can precipitate suboptimal model performances, while the latter can, either inadvertently or with malice aforethought, result in the exposure of sensitive training datasets.

Moreover, while the multifaceted functionalities of FL systems are tailored to accommodate diverse user prerequisites, this very versatility augments the complexity of the system, potentially exacerbating its vulnerability footprint.

\subsubsection{Risk Management}

Risk management within computational domains encapsulates a triad of cardinal components: assessment, control, and surveillance \cite{Laura0FISMA}.
Assessment: This phase is devoted to the recognition and quantification of potential risks, encompassing both system vulnerabilities and adversarial intrusions. It emphasizes discerning the nature of these threats and prognosticating their potential ramifications on the computational infrastructure.
Control: Rooted in the analytics of the assessment phase, control prioritizes the deployment of mechanisms to alleviate identified threats. Its objectives are manifold: to meticulously scrutinize potential threats, curtail prospective damages, and to manifest robust security protocols.
Surveillance: This perpetual process underscores real-time oversight and feedback mechanisms to gauge the efficacy of the implemented security apparatus.

Within the ambit of FL, preempting and mitigating risks are of paramount importance to uphold an impeccable security standard \cite{2020SecureAF}. Two salient strategies emerge to fortify security.
\textbf{Environment Fortification:} The inception of a fortified computational milieu, exemplified by a Trusted Execution Environment (TEE) \cite{hunt2018ryoan}, offers a bulwark against many extant risks.
\textbf{Dynamic Risk Surveillance:} Constructing a dedicated surveillance module can yield dividends in proactively identifying and mitigating security anomalies. Such a module functions ceaselessly, scrutinizing the FL framework for aberrations, thereby enabling the formulation and deployment of nimble preventive stratagems.

%\textbf{Trusted Execution Environment (TEE): A Hardware-centric Privacy Paradigm.} 

TEE emerges as a hardware-centric approach to privacy-preserving computations. This solution empowers remote users to execute computational tasks on a device, effectively cloaking the intricacies of the computations from the hardware fabricator. Leveraging specialized CPU registers and ensuring memory isolation or encryption, TEE fosters a sanctuary for secure computations \cite{hunt2018ryoan}. As the trustworthiness of the TEE is ascertained, the platform extrapolates a cryptographically secure interaction milieu between devices, potentially amplifying the efficacy of collaborative endeavors like FL \cite{chen2020training, mo2021ppfl}. Within the FL context, TEEs are perceived as potent aggregation centers, streamlining parameter consolidation \cite{zhang2021shufflefl}. A nascent exploration into blockchain-augmented TEEs hints at an inviolable approach, rendering local model manipulations futile \cite{kalapaaking2022}. Additionally, the integration of a TEE-oriented proxy component can fortify the confidentiality of FL participants, ensuring the integrity of updates dispatched to servers \cite{2021MixNNPO}. Notwithstanding its commendable efficacy, the resilience of the TEE is tethered to its foundational hardware schema \cite{2020Zhao}, rendering it potentially susceptible to specific adversarial interventions, such as data poisoning exploits \cite{2021Mondal}.

%\textbf{Risk Surveillance and Governance in FL Architectures}

The overarching paradigm of risk surveillance and governance encompasses meticulous oversight of the entire training lifecycle, spanning risk detection, quantification, assessment, and redressal. The quintessential objective herein is to refine the security stratagem and ascertain alignment with stipulated security benchmarks. Through real-time oversight, potential adversarial vectors and insufficiencies in security postures are unveiled, thus paving the way for timely recalibration of protective measures.

In FL, a dedicated surveillance component is pivotal to oversee the process of training and system dynamics. Such a component, while granting developers granular feedback, simultaneously facilitates risk attenuation. An example of this ideology is the FATE framework, which boasts the FATE-Board module. This visual interface meticulously chronicles task execution trajectories and model performance metrics \cite{FATE}, engendering an enhanced oversight mechanism for FL systems. This operational transparency aids risk elucidation and sanctions dynamic recalibrations of the security posture.

\subsection{Malicious Attacks on FL}
\label{sec:Malicious}

Within the expansive landscape of computational systems, certain vulnerabilities possess the propensity to critically perturb both the functional and computational efficiency of a system \cite{2019Privacy}. Astute adversaries, having discerned these susceptibilities, are equipped to orchestrate intricate intrusions that imperil system integrity \cite{MOTHUKURI2021619, 2021Challenges, 2020Tian, BLANCOJUSTICIA2021}. Compounding these complexities is the presence of malevolent participants, who, embedded within the network of the system, can initiate a myriad of sophisticated threats \cite{2021Privacy, 2020Privacy}. These range from poisoning offenses \cite{2020Backdoor} and Byzantine stratagems \cite{2018DRACO, 2021So}, to intricate inference attacks \cite{2020From}.

In the ensuing discourse, we delve into a meticulous examination of prevalent adversarial modalities afflicting FL architectures: poisoning incursions, adversarial maneuvers, and inference transgressions. Please note that the onslaughts of both poisoning and adversarial character predominantly manifest during the intricate training phase of FL.

\subsubsection{Poisoning Attacks in FL}

Poisoning attacks, insidious by nature, are orchestrated through the intentional adulteration of training datasets, aiming predominantly at the degradation of model efficacy. A dichotomous classification bifurcates these attacks into ``data poisoning'' and ``model poisoning'' \cite{2021Wang}.

\textit{Data Poisoning}: This strand comprises two predominant subsets. The first, ``label-flip`` attacks, involves the surreptitious manipulation of label metadata within the training set, inducing significant deviations in model targets and consequent attenuation of accuracy \cite{2015Support}. The latter, termed ``clean-label`` attacks, are characterized by the subtle modification of the training dataset, or the infusion of strategically erroneous data, inevitably leading to the degradation of model precision.

\textit{Model Poisoning}: This paradigm hinges on the direct manipulation of model parameters or its architecture. Canonical examples include backdoor attacks, wherein the global performance of the model remains ostensibly unaffected, while yielding skewed results for specific input domains \cite{2017BadNets, 2018Chen, 2022Schulth}.

FL is acutely susceptible to poisoning attacks, given its unique architectural tenets. The inherent data heterogeneity in FL, often manifesting as non-Independent and Identically Distributed (non-IID) data, culminates in localized model variations across devices \cite{2021Duan, 2021Wu, 2022Xiong}. Further exacerbating the vulnerability is the systemic design where the central server remains detached from the granularities of the training process, rendering the validation of device-originated updates a formidable challenge \cite{Fang2019LocalMP, 2021PoisonGAN}. 
In addition, recent works have presented federated unlearning \cite{che2023fast} as a solution to the challenges posed by data heterogeneity in FL \cite{wang2023federated}. Through the targeted removal of particular local data, these methods interfere with the specific knowledge that the system derives from such data, thereby enhancing local data privacy \cite{zhu2023heterogeneous}.
Lastly, the sheer voluminosity of participating devices in FL amplifies the complexity of anomaly detection, often obscuring the malicious entities operating within this extensive landscape \cite{2018Analyzing, 2018Mitigating}.

\subsubsection{Adversarial Attacks in FL}

Adversarial attacks, characterized by the strategic infusion of minuscule alterations within training datasets, are tailored to mislead target models \cite{2018FeatureDF}. Remarkably, these diminutive alterations can precipitate significant aberrations, including misclassifications. Grounded on the extent of model knowledge available to the attacker, adversarial attacks bifurcate into white-box, gray-box, and black-box categories.

\begin{table*}[t]
  \caption{Defense Methods of Poisoning Attack}
  \label{tab:Poisoningattack}
  \centering
  \scalebox{0.7}{
    \renewcommand{\arraystretch}{1.25}
	\begin{tabular}{|c|c|c|c|m{7cm}|}
		\hline
		Attack Types & Defense Type & Algorithm & Year & Description \\
    \hline\hline
\multirow{6}{*}{{Data Poisoning}}  & \multirow{6}{*}{{Data Preprocessing}} & AUROR \cite{2016AUROR} & 2016 & The method uses a clustering algorithm to detect anomalous training data. \\
\cline{3-5}   & & Carlini et al., \cite{2018The} & 2018 &  This method is based on statistical methods for preprocessing training data. \\
\cline{3-5}   &  & Zhang et al.  b\cite{2021SAFELearning} & 2021 & The security measures are based on a random selection of clients to reduce the impact of malicious participants.  \\
\cline{3-5}   & & Client-Side Detection \cite{Zhao2021} & 2021 & The security mechanism randomly selects a group of clients to evaluate the model, and the server adjusts the weight according to the results. \\
\cline{3-5}   & & SEAR  \cite{2021SEAR} & 2021 & The method is based on data sampling, and it can effectively detect Byzantine faults without reducing the performance of the model. \\
\cline{3-5}   & & FLDebugger \cite{2022Debugging} & 2022 & The framework includes a debugging module to reduce the impact of incorrect training data.  \\
 \hline
 
\multirow{12}{*}{{Molel Poisoning}} & \multirow{6}{*}{{Anomaly Detection}} & Sniper \cite{Cao2019} & 2019 & The method uses Euclidean Distance between local models to detect malicious participants. \\
\cline{3-5}    &  & FoolsGold \cite{Fung2020} & 2020 & The method uses Cosine Similarity between device updates in order to detect malicious participants.\\
\cline{3-5}   &   & PEFL \cite{2021Enhanced} & 2021 & The method compares the encrypted gradient vector with the median vector distance in order to detect malicious participants. \\
\cline{3-5}   &   & CONTRA\cite{2021CONTRADA} & 2021 & The method uses Cosine Similarity to determine the credibility of local model parameters in each round.   \\
\cline{3-5}   &   & Chen et al., \cite{2021Chen} & 2021 &  The method uses two anomaly metrics, namely the relative distance and the convergence measure to detect anomalies. \\
\cline{3-5}    &  & Ma et al., \cite{2022Ma} & 2022 & The method evaluates the distance between two encrypted gradients in order to detect malicious updates. \\

\cline{2-5} & \multirow{6}{*}{{Model Robustness}} & SFPA \cite{2020Pocket} & 2020 & The framework is based on multi-key computation for addressing the leakage of a single key. \\
\cline{3-5}   & & FL-Block \cite{2020FLBlock} & 2020 & This framework uses blockchain technology to prevent poisoning attacks. \\
\cline{3-5}   &  &  Hashgraph-based Method \cite{2021Hashgraph} & 2021 & The method uses hash graphs to protect user privacy and detect malicious participants. \\
\cline{3-5}   & & RoFL \cite{2021RoFL} & 2021 & This security mechanism is based on zero-knowledge proof, and enhances security aggregation. \\
\cline{3-5}  & & Turbo \cite{2021Turbo} & 2021 & The method employs a multi-group circular strategy for model aggregation to increase system robustness. \\
\cline{3-5}  & & SparseFed \cite{2022SparseFed} & 2022 & This method is based on gradient clipping in order to protect against poisoning attacks. \\
   \hline
	\end{tabular}
	}
\normalsize
\end{table*}

\textit{White-box Attacks}: Predicated on the presumption that the attacker has comprehensive knowledge about the architecture and parameters of the target model, these attacks fabricate adversarial examples to directly misguide the model. Seminal algorithms, such as the BFGS attack \cite{2013Intriguing}, ascertain misclassifications by discerning the minimal loss function perturbations. Techniques like FGSM \cite{goodfellow2014} and its iterative counterpart, I-FGSM \cite{tramer2017space}, employ gradient step computations to generate these adversarial exemplars. The DeepFool algorithm \cite{2016deepfool} and the JSMA method \cite{2016limitations} further delineate efficient adversarial example creation by computing minimal requisite perturbations and exploiting forward propagation derivatives, respectively.

\textit{Gray-box Attacks}: Operating under partial knowledge of the target model, such as its structure or training data, these attacks predicate on a surrogate model, trained using the extant model knowledge. This surrogate assists in the generation of adversarial examples \cite{2021MetaAttack}.

\textit{Black-box Attacks}: These attacks, characterized by a lack of intrinsic knowledge on the part of the attacker about the target model, employ gradient estimations to craft adversarial examples \cite{2017zoo}. Distinctive methodologies encompass evolutionary algorithms \cite{2019yet}, meta-learning paradigms \cite{2019query}, and Bayesian optimization techniques \cite{2019bayesopt}. Notably, the transferability of adversarial examples across models renders black-box attacks particularly potent, facilitating the potential compromise of unrelated models \cite{2016Transferability, QianHuang2019}.

In the context of FL, the decentralized nature and dynamic participation render it susceptible to adversarial attacks. The omnipresent global model parameters render white-box and gray-box attacks more plausible. Conversely, black-box attacks, albeit hindered by prolonged query durations and diminished efficacy, present a more latent threat, given the systemic predomination of the global model in FL.

\subsubsection{Inference Attacks in FL}

Inference attacks in machine learning pivot on exploiting the model to unveil obscured training data or discern sensitive attributes. Predicated on the kind of information being pursued, these attacks bifurcate into property inference attacks and membership inference attacks.

\begin{table*}[t]
  \caption{Defense Methods Of Adversarial Attack}
  \label{tab:Adverattack}
  \centering
  \scalebox{0.7}{
    \renewcommand{\arraystretch}{1.25}
	\begin{tabular}{|c|c|m{11cm}|c|}
		\hline
		Algorithm & Year & Description & Defense Type \\
    \hline\hline
  HGD \cite{liao2018defense} & 2018 & The method adds denoising techniques to the model to reduce the error introduced by adversarial samples.  & \multirow{6}{*}{{Complete}}\\
\cline{1-3}   AGKD-BML \cite{wang2021agkd} & 2021 & The framework uses knowledge distillation techniques \cite{dong2022elastic,liu2023large} to develop a teacher-student model. Attentional knowledge is extracted from the teacher model and it is transferred to the student model to increase model accuracy by focusing on the correct region. &  \\
\cline{1-3}    FDA$^{3}$ \cite{song2020fda} & 2020 & The framework includes an attack monitor module that is used to record attack information, and conduct adversarial retraining. & \\
\cline{1-3}    DNE \cite{zhou2020defense} & 2020 & This method minimizes adversarial perturbations by randomly sampling and replacing words.& \\
\cline{1-3}    GDMP \cite{wang2022guided} & 2022 & The method uses diffusion models as a preprocessor to denoise the input. & \\
\cline{1-3}    RSLAD \cite{zi2021revisiting} & 2021 & The framework uses knowledge distillation techniques to develop a teacher-student model to learn new labels associated with the new data, and then utilizes the student model to ensure security of the new labels. & \\
    \hline
    SafetyNet \cite{lu2017safetynet} & 2017 & The method detects perturbations in an image. & \multirow{4}{*}{{Detection}} \\
\cline{1-3}    MagNet \cite{meng2017magnet} & 2017 & The method uses one or more external detectors to classify an input image as adversarial or clean. & \\
\cline{1-3}    Feature squeezing \cite{xu2017feature} & 2017 & The method compares the prediction results between the original image and squeezed images. & \\
\cline{1-3}    ZeKoC \cite{2021Zero} & 2017 & The method uses a clustering algorithm to detect malicious participants. & \\
   \hline
	\end{tabular}
	}
\normalsize
\end{table*}

\textit{Property Inference Attacks}: These attacks endeavor to deduce concealed or fragmented attributes leveraging overtly available properties or data distributions \cite{2021HCN}. As an exemplar, within a recommendation system paradigm, a malevolent entity could discern pivotal attributes such as age, gender, or even deeper personal nuances based on the evident model outputs—like frequent purchase patterns or items earmarked as intriguing.

\textit{Membership Inference Attacks}: Operating on the juxtaposition of the target model and a data exemplar, these attacks endeavor to ascertain the affiliation of the sample with the training set of the model, probing if it was pivotal during the training epoch \cite{2021Truex}.

FL manifests as a decentralized learning paradigm where clients retain their data and disseminate specific parameters, predominantly gradients, to a centralized entity for synergistic model training. Notwithstanding its decentralized veneer, FL is inherently susceptible to nefarious incursions, particularly from malicious system participants. These malevolent entities may be embodied as either a client or a server.

\textit{Server-Side Adversaries}: A compromised server, privy to intricate nuances of the local training model (spanning model architecture, client identities, and gradients), possesses the prowess to decipher concealed client attributes, a manifestation of property inference attacks \cite{2017PPDL}. Such entities may also employ Generative Adversarial Networks (GANs) to reverse-engineer client updates, endeavoring to recreate the original training data \cite{2019Wang,2020Song}.

\textit{Client-Side Adversaries}: These malefactors, leveraging the periodic updates from specific clients, can deduce intricate and sensitive client data \cite{2019exploiting}. The adversarial landscape also encompasses the perpetration of continuous spurious data injections, convoluting the training regime. Such incursions, colloquially termed poisoning attacks, can precipitate inadvertent divulgence of sensitive attributes \cite{xin2020private,2017Hitaj}.

\subsection{Defensive Paradigms in FL Systems: Approaches and Classifications}

In the intricate landscape of FL systems, malevolent incursions manifest in multifarious guises, necessitating a diverse repertoire of defense mechanisms. Distinct defense stratagems are earmarked based on the potential location of the attack, i.e., on device or server side, while overarching security measures often pivot on systemic robustness \cite{2020Chuan}. The threat landscape in FL encompasses potential adversaries domiciled either within client devices or centralized servers.

A salient stratagem to bolster security, especially against compromised servers, amalgamates the principles of distributed architectures combined with advanced encryption methodologies. From a broader perspective, defense mechanisms in FL can be dichotomized into two predominant categories: proactive and reactive defense paradigms \cite{MOTHUKURI2021619}. \textit{Proactive Defense}: As the nomenclature suggests, proactive defense initiatives prognosticate potential threat vectors, orchestrating and deploying defense mechanisms \textit{a priori} to forestall the actualization of anticipated attacks. \textit{Reactive Defense}: Reactive defense mechanisms are reactionary, coming to the fore post facto, in the aftermath of an attack detection, working towards mitigation and remediation.

%In this section, we detail the defense approaches.

\subsubsection{Countermeasures for Poisoning Attacks}

FL systems have witnessed the proliferation of diverse poisoning attacks during the training phase, prompting the genesis of specialized countermeasures. This discourse delineates these defensive mechanisms, specifically addressing data and model poisoning attacks. A categorization complemented by paradigmatic defenses against these attacks, informed by extant literature, is encapsulated in Table \ref{tab:Poisoningattack}.

\textbf{Defensive Strategies for Data Poisoning Attacks:}
The fulcrum of defenses against data poisoning pivots on two crucial pillars: the reliability and integrity of the training data \cite{2021Girgis}. Reliability underscores the authenticity of training data, and integrity ensures alignment with global data distributions and prescribed formats. Notably, defense measures are discerned from a data-centric protective lens. Methods like random sampling of training data \cite{Blanchard2017, Mhamdi2018TheHV} and anomaly detection have demonstrated efficacy. However, in a multi-device FL milieu, challenges persist, particularly in the selection of honest computing nodes. Techniques ranging from rule-based \cite{2021SAFELearning, Zhao2021} to sampling-based data selection \cite{2021SEAR, 2021SAFELearning}, and data purification \cite{2018The, wang2022guided, wu2022guided} have emerged in response. The AUROR framework, utilizing the Euclidean distance metric, exemplifies how poisoned data can be identified and culled \cite{2016AUROR}. Nonetheless, these methodologies are not bereft of challenges, with the pursuit of an optimal balance between performance and security remaining elusive \cite{2021Girgis}.

\textbf{Defensive Strategies for Model Poisoning Attacks:}
To handle model poisoning, the emphasis shifts towards discerning malicious entities or spurious updates during the aggregation phase. Model anomaly detection, typified by techniques such as measuring the Euclidean distance among updates, emerges as a potent tool for isolating aberrant models \cite{2021CONTRADA, 2021Chen, Cao2019, 2022Ma, 2021Enhanced}. Another avenue is assessing the contribution of each device \cite{Fung2020}. Enhancing the inherent robustness of models offers another bulwark against poisoning. This can be achieved through myriad strategies: pruning the model \cite{2021Hashgraph, 2022SparseFed}, fortifying with encryption \cite{2021RoFL, 2020Pocket, 2021Turbo}, or the integration of blockchain for integrity assurance \cite{2020FLBlock}. Retraining poisoned models has also been proposed as an effective countermeasure \cite{song2020fda}.

\begin{table*}[t]
  \caption{Defense Methods Of Inference Attack}
  \label{tab:Infattack}
  \centering
  \scalebox{0.7}{
    \renewcommand{\arraystretch}{1.25}
	\begin{tabular}{|c|c|m{9cm}|c|}
		\hline
		Algorithm & Year & Description & Defense Type \\
    \hline\hline
    VerifyNet \cite{2020VerifyNet} & 2020 & The framework employs a double-masking protocol to ensure the confidentiality of the gradient. & \multirow{3}{*}{{Encryption}}\\
\cline{1-3}    Privacy-preserving entity resolution \cite{2017Private} & 2017 & The method use HE to preserve privacy in VFL. & \\
\cline{1-3}    FedMONN \cite{2020FedMONN} & 2020 & The framework with an encoder-decoder architecture. While the client encrypting updates, the server decrypts them, and protects the privacy of updates during transmission. &\\
    \hline
   MemGuard \cite{2019MemGuardDA} & 2019 & The method can mislead attackers by adding noise to each confidence score vector predicted by the target classifier. & \multirow{5}{*}{{Obfuscation}} \\
\cline{1-3}    Gradient noise \cite{2021GradientIA} & 2021 & The method inserts Gaussian noise to gradients to defend against attacks. &  \\
\cline{1-3}    Client-level DP algorithm \cite{2017Differentially} & 2017 & The method randomly selects some devices to compute perturbation values randomly, and updates the central model using these perturbations. & \\
\cline{1-3}   CoAE \cite{2022DefendingLI} & 2022 & The method replays the label of the training set to defend against inference attacks. & \\
\cline{1-3}    Digestive Neural Networks(DNN)\cite{2021DigestiveNN} & 2021 & The method uses DNN to transform the data of the training set in order to protect privacy. &  \\
    \hline
   Privacy-Preserving FL\cite{2019HybridTruex} & 2019 & The method combines DP with MPC and reduces noise injection as the number of parties increases. & \multirow{3}{*}{{Hybrid Method}}\\
\cline{1-3}    Extensions based on Sharemind \cite{2015Combining} & 2015 & The method combines DP with Sharemind (a MPC platform) for the purpose of protecting the privacy of data providers and individuals. & \\
\cline{1-3}    SGD based method \cite{2019Meng} & 2019 & The method combines HE and DP to ensure data privacy and reduce communication costs. & \\
   \hline
	\end{tabular}
	}
\normalsize
\end{table*}

\subsubsection{Countermeasures Against Adversarial Attacks}

In FL, adversarial attack defenses can strengthen the robustness of the system by incorporating more data samples, e.g., adversarial training \cite{2019feature}, data augmentation or compression \cite{2020Sattler}. In addition, the defense capability of the system can also be improved by anomaly detection, e.g., anomaly client detection \cite{2021Zero}, adversarial examples detection \cite{liao2018defense}. In general, defense methods can be divided into complete defense and detection-only defense \cite{Akhtar2018}. A summarization of prevalent defense techniques is available in Table \ref{tab:Adverattack}.

\textbf{Complete Defense Mechanisms:}
Complete defense mechanisms aim to mitigate the influences of adversarial examples by ensuring accurate classification during the training phase. These countermeasures encompass strategies such as:
\begin{itemize}
\item Introducing an attack monitoring module for continuous vigilance and iterative model refinement \cite{song2020fda}.
\item Deploying knowledge distillation for accurate classification of adversarial samples via teacher-student model configurations \cite{liu2022large, wang2021agkd, zi2021revisiting}.
\item Employing data augmentation techniques, wherein filters such as noise addition are integrated to restrain inaccuracies instigated by adversarial inputs \cite{gowal2021improving, liao2018defense}.
\item Utilizing the diffusion model, where a clean datum undergoes controlled contamination and subsequent iterative noise removal, offering a dual-process defense against adversarial samples \cite{2022diffusion, wang2022guided, wu2022guided, choi2022perception}. However, this approach comes with computational complexity, impinging on training efficiency \cite{nie2022diffusion}.
\end{itemize}

\textbf{Detection-only Defense Mechanisms:}
Detection-only defenses prioritize identifying adversarial samples without necessarily correcting them. Representative methodologies encompass:
\begin{itemize}
\item Leveraging data disturbances to pinpoint adversarial inputs \cite{lu2017safetynet}.
\item Utilizing compression techniques to gauge data variations pre and post-compression, thereby discerning adversarial nuances \cite{xu2017feature}.
\item Incorporating external detectors within systems to continuously monitor and detect adversarial entities \cite{meng2017magnet}.
\end{itemize}

Modern defense mechanisms grapple with the equipoise between efficiency and efficacy. Adversarial training, albeit competent, remains computationally taxing and does not assure comprehensive coverage against all adversarial samples, exposing its limitations \cite{bai2021recent, de2020survey}. On the other hand, seemingly facile techniques like randomization and denoising, despite their easy deployment, do not guarantee consistent defense efficacy \cite{vargas2019robustness}.

\subsubsection{Countermeasures Against Inference Attacks}

The intricacies of FL systems include safeguarding a spectrum of sensitive information: training data, model constituents (algorithms and parameters), and resultant outputs. A breach in any of these facets can profoundly undermine the integrity of the system \cite{2019Quantification, 2019Deep}. The plausible incorporation of malevolent entities further accentuates the vulnerability to information exploits \cite{MOTHUKURI2021619}. A synthesized overview of prominent defense strategies against inference attacks is delineated in Table \ref{tab:Infattack}. Subsequently, we delve into the nuances of these defense paradigms.

\textbf{Encryption-centered Techniques:}
Integral techniques like HE and Secret Sharing (SS) stand at the forefront of encryption practices. The prowess of HE lies in its ability to conduct operations on encrypted data directly, ensuring privacy throughout the computational process \cite{2017Private,fang2021privacy}. SS, on the other hand, involves the dissemination of key shares across multiple entities, enabling decryption only by authorized coalitions \cite{2017Bonawitz}. Yet, the dual adoption of encryption algorithms, while enhancing data sanctity, can inadvertently increase the computational and communicational overheads in FL systems. The quintessential challenge manifests as striking a balance between maintaining efficient training and assuring model accuracy \cite{fang2021privacy, zhang2020batchcrypt, ma2022privacy}. The encoder-decoder architecture offers a potential solution by allowing devices to perform data encryption and servers to undertake decryption \cite{2020FedMONN}.

\textbf{Obfuscation-centered Techniques:}
DP remains a pivotal technique, adding deliberate obfuscation to the data or specific features, thereby ensuring that third-party entities are unable to distill individual raw data from device-transferred information \cite{asoodeh2021three, 2022DefendingAM,2019MemGuardDA}. Current obfuscation techniques predominantly involve the infusion of Laplacian or Gaussian noise \cite{2021GradientIA,2020Wei,liang2020}. Augmenting training data or manipulating label information further intensifies data obscurity \cite{2022DefendingLI, 2021DigestiveNN}. Nonetheless, the dilemma persists: excessive noise infusion may degrade model precision, while sparse noise addition might inadvertently expose training data \cite{2017Differentially, ren2022grnn}.

\textbf{Hybrid Techniques:}
Considering the constraints intrinsic to pure encryption or obfuscation techniques, hybrid methods like MPC emerge as a promising alternative \cite{2015Combining}. Contemporary MPC protocols, coupled with DP technologies, aim to both preserve data privacy and curtail communicational overheads \cite{2019Meng}. Modulated noise injection further refines model accuracy \cite{2019HybridTruex}.

As the narrative underscores, while there exists a plethora of defense mechanisms, achieving an optimal trade-off between model accuracy and robust defense against inference attacks remains a persistent challenge in the realm of FL.

\section{Application of FL}
\label{sec:FLApplication}

Users and policymakers are increasingly aware of the importance of data security and privacy within FL systems. This has led to a surge in research on privacy-protection measures, and data access is being scrutinized more closely.

As FL continues to evolve and be adopted across various industries, it is important to recognize and address the potential vulnerabilities and risks inherent in these applications. By doing so, we can design more robust and secure FL systems that can withstand specific attacks targeting these applications. FL is currently applied in various areas, such as healthcare \cite{2020healthcare, 2020Secure,2020Decentralized}, IoT \cite{2021IoTSurvey}, autonomous vehicles \cite{2020Zhang, 2021Yuan}, finance \cite{schreyer2022}, wireless technology \cite{2021Mohamed}, and recommendation systems \cite{2021Recommender}.

\subsection{Healthcare}

In healthcare, FL is used to allow medical institutions to train models independently using their local data, ensuring patient privacy \cite{2014Systematic, 2022Healthcare}. However, a major vulnerability arises due to the sensitive nature of healthcare data. An attack could potentially disrupt the learning process, inject malicious models, or leak sensitive patient information \cite{2019Federated}. Techniques such as DP and MPC can help mitigate these risks, but they need to be carefully incorporated to maintain the balance between privacy and model performance  \cite{2021Dopamine}. 

\subsection{Finance}

In the finance industry, FL allows institutions to share insights without disclosing sensitive data, thus maintaining privacy  \cite{schreyer2022}. Yet, FL in finance also presents an attractive target for adversarial attacks aiming to manipulate the consensus model for illicit financial gain, or to compromise financial data confidentiality \cite{basu2021, imteaj2022}. Techniques such as robust aggregation and Byzantine fault tolerance can help build resilience against these attacks, but more research is needed to optimize these defenses for finance-specific scenarios \cite{byrd2020}.

\subsection{Wireless Communications}

\chen{FL in wireless communications technology aims to preserve data privacy by training machine learning models in a decentralized manner \cite{2021Mohamed}. However, this decentralization can also introduce vulnerabilities as attackers may exploit insecure communication channels to extract sensitive information or inject adversarial models \cite{2021Wireless, 2021ChenMingzhe}. Additionally, jamming attacks can pose a serious threat to the stability and security of wireless communication networks, especially in dense network environments where attackers can use jamming signals to interfere with data transmission, reducing the effectiveness of FL model training and prediction accuracy \cite{Houda2024Jamming}. Solutions such as secure communication protocols and robust FL algorithms are crucial, but challenges remain in designing these solutions to be both effective and efficient in the wireless environment \cite{2020Wireless, 2022Liu}.}

\subsection{Smart Transportation}

\chen{The application of FL in smart transportation presents unique challenges \cite{samarakoon2019}. For instance, attackers may target the decentralized nature of the data, aiming to compromise data related to transportation systems or manipulate the consensus model, thereby causing safety hazards. Furthermore, the real-time and mobile nature of smart transportation systems imposes additional constraints on defenses \cite{Chen2021BDFL,Pokhrel2020Autonomous}. Techniques such as on-the-fly data encryption and real-time anomaly detection can help secure FL in this domain, but their effectiveness and impact on system performance need further investigation.}

\subsection{Recommendation Systems}

Recommendation systems greatly benefit from FL as it allows for personalized recommendations while protecting user data \cite{ammad2019federated,wu2021hierarchical}. However, the personalized nature of these systems also creates potential vulnerabilities \cite{narayanan2008robust}. An attacker might infer sensitive user information from model updates, or manipulate the learning process to affect the recommendations. Privacy-enhancing techniques such as DP and HE can provide protection, but how to effectively integrate these techniques in FL-based recommendation systems is an ongoing research topic \cite{lin2020fedrec, liang2021fedrec, lin2020meta}.

\subsection{Smart Cities}

Smart cities, encompassing traffic management, environmental monitoring, and utility services, have become an emerging application for FL. The advantage of using FL lies in its ability to process and learn from vast amounts of data generated by various sensors and devices in the city while preserving privacy \cite{jiang2020federated, hanjri2023federated}. However, this interconnectivity and heterogeneity can expose the system to targeted attacks such as data poisoning or model inversion attacks \cite{ALHUTHAIFI2023833}. Ensuring data integrity, enforcing MPC protocols, and integrating privacy-preserving techniques like DP are vital for mitigating these security risks. 
\chen{In addition, recent research has enhanced system resilience against local model manipulation by integrating blockchain technology, significantly bolstering the capabilities of intrusion detection systems to defend against sophisticated attacks \cite{Houda2023}.}

\subsection{Generative AI in FL}

\chen{The rise of generative AI, especially Generative Adversarial Networks (GANs), offers new possibilities in enhancing the defense mechanisms within FL systems. GANs, by their design, can play a pivotal role in identifying and mitigating data poisoning attacks, a significant threat to FL environments \cite{psychogyios2023gan}. For example, GANs can be used to generate synthetic data that mimics the characteristics of poisoned datasets, which then helps in training FL models to recognize and reject malicious inputs effectively \cite{laykaviriyakul2023collaborative}. By incorporating GAN-based strategies, researchers can advance the security measures necessary to protect decentralized learning processes from evolving threats \cite{zhang2020exploiting}.}

\section{Open Challenges and Future Directions}
\label{sec:challenges}

FL is a distributed learning technology that provides privacy protection. Despite FL enabling devices to train a model without sharing raw data, several security and privacy concerns persist. This section explores various challenges and future directions of FL systems.

\subsection{Reliability and Security of Big Models}

FL systems face unique challenges related to the size and complexity of deployed models. The number and capacity of devices involved in FL systems can limit the application of large-scale models \cite{ro2022scaling}. As computational resources and data volume increase, models scale up, often necessitating a larger amount of raw data for training. Consequently, this leads to an increase in the number of model parameters and a more complex structure \cite{yuan2022roadmap}. Examples of such large models include DALL$\cdot$E \cite{dalle} and DALL$\cdot$E2 \cite{dalle2} for text-to-image generation, and pre-trained language models such as Bert \cite{clark2020electra}, GPT-3 \cite{2020gpt}, and XLNet \cite{2019xlnet} designed for NLP tasks. The WuDao model \cite{2022wudaomm}, boasting 1.75 trillion parameters, is an example of scaling up parameters to enhance performance across a broad spectrum of machine learning problems \cite{liu2022dive}.
Large models, despite their performance advantages, introduce a higher degree of vulnerability \cite{carlini2021}. For instance, large datasets can obscure maliciously introduced poisoning data, and existing defense methods might encounter scalability problems when dealing with larger models \cite{yuan2022roadmap}. Furthermore, these models are more likely to inadvertently retain sensitive raw data \cite{tirumala2022, carlini2019secret}, posing considerable privacy risks to the FL system \cite{carlini2022privacy}. Attacks such as inference attacks or data reconstruction attacks present severe privacy risks to large models \cite{yuan2022roadmap,liu2024fisher,che2023federated}.

Given these risks, it becomes imperative to thoroughly understand the security and privacy implications of large-scale models within FL systems. Furthermore, current defense methods must be critically evaluated and developed further to adequately address the risks associated with these large models.
While designing these defense methods, a crucial consideration is to maintain the accuracy of the models. This balance can be achieved by integrating privacy-preserving techniques such as DP or HE in the learning process, adopting robust aggregation methods to limit the impact of malicious updates, and employing model pruning or compression strategies to maintain model accuracy while reducing model size and complexity. Ultimately, our goal is to devise defense strategies that effectively address the security and privacy concerns of large-scale models within FL systems, without compromising their accuracy.

\subsection{Dynamic Adaptive Defense Techniques}

FL frameworks enable collaborative training across multiple institutions while ensuring privacy, data security, and compliance with regulatory requirements. These systems involve a wide array of technologies, including infrastructure frameworks, data distribution and storage, algorithms, communication, and deployment. However, storing training data locally in FL systems presents a significant risk of data leakage or exploitation by malicious entities. Moreover, the heterogeneity of the data can make it challenging to identify such attacks.
Additionally, vulnerabilities in system environments, such as clusters or cloud platforms, may expose FL systems to security threats like Denial of Service (DoS) attacks and unauthorized access \cite{Berman2019ASO}. At present, no reliable method exists for quantitatively assessing the risk associated with FL systems. 
While some static defense measures have been used to protect FL systems, these often lack effectiveness and adaptability in the face of changing threat landscapes. Consequently, we propose the use of dynamic adaptive defense techniques. These are security measures that monitor system status in real-time and adapt their defenses based on identified threats and the specific application scenarios.

Dynamic adaptive defense techniques can help make FL systems more robust against evolving threats. For example, traditional risk monitoring techniques can be enhanced by incorporating real-time threat detection and adaptive response mechanisms. An application of such a dynamic adaptive defense technique might involve monitoring the model updates from each participant in an FL system and dynamically adjusting the response of the system if suspicious activity is detected, such as outlier updates that might indicate an attack.
Additionally, the system could adapt its defenses based on the context of the application. For instance, an FL system used in a healthcare setting might require stricter privacy controls and might employ techniques like DP to dynamically adjust the level of noise added to the data based on the sensitivity of the data being processed.
In summary, dynamic adaptive defense techniques can provide a more flexible and robust defense against security threats to FL systems, and further research and development of these techniques should be a priority.

\subsection{Lightweight Cryptography}

FL offers a unique framework for institutions to collectively train a model without needing to share data directly, ensuring compliance with legal and regulatory requirements. However, existing FL approaches are not exempt from various security and privacy concerns. A malicious participant could potentially extract sensitive information from the training results, which could have an adverse effect on specific targets. Similarly, malicious servers may infer training data from client updates, which could compromise the integrity of the global model.
Current solutions aim to enhance privacy by integrating techniques such as DP, MPC, or HE \cite{2017Differentially, 2021Su, 2020Wei}. Of these, MPC is growing in acceptance as a method to boost model performance while maintaining security \cite{2019HybridAlpha}. In addition, the emergence of blockchain technology has supplemented FL by offering a decentralized, traceable, and recordable solution.
Despite their benefits, these measures can potentially reduce model accuracy or efficiency. For instance, MPC can be resource-intensive, pushing participants to lower the security level to decrease data transmission costs and improve training effectiveness \cite{2020Communication, 2016FederatedLS}.
In this context, "low loss" refers to minimizing any reduction in model performance, including factors like accuracy, precision, and recall, when incorporating cryptographic techniques. The term "high efficiency" is defined as the capability to perform computations and data transmission swiftly with minimal resource usage.
Considering these definitions, the need arises for robust, privacy-preserving, and lightweight technologies that can ensure minimal loss in model performance and high computational and communicational efficiency. By developing and implementing such techniques, we can achieve a better balance between security, privacy, and performance in FL systems.

\subsection{Performance Efficiency}

In FL systems, cryptographic technologies are often used to enhance security and maintain user privacy. Nevertheless, the inherent complexity of these technologies can compromise the overall system efficiency. This is seen in many applications where cryptographic techniques are used extensively, leading to a substantial increase in computation time or other resources.

Several strategies have been explored to mitigate this effect, such as the use of model compression techniques \cite{2022Aggregation} or minimizing communication costs \cite{2022SwiftAgg}, aiming to strike a delicate balance between system security and computational efficiency.
To illustrate this trade-off, consider an FL system that employs advanced encryption for secure communication. While this method effectively preserves data privacy, it also incurs significant computational overhead, resulting in decreased overall performance. Conversely, a less secure but computationally lightweight encryption method might enhance system performance but at the cost of reduced security.
The challenge in setting a quantitative standard to balance security and efficiency arises from the multitude of factors that contribute to both. System security, for example, can be influenced by the strength of encryption, data anonymization techniques, and robustness against attacks, among others. On the other hand, computational efficiency may depend on factors like data volume, model complexity, and network bandwidth.

While no solution fits all sizes for this issue, potential strategies might involve developing adaptive cryptographic techniques that adjust their level of security based on the computational capacity of the system or using a multi-objective optimization approach that balances both security and efficiency metrics. However, both of these approaches require extensive research and careful implementation, highlighting the complexity of this challenge \cite{2021DP, 2020Efficient}.

\section{Conclusion}
\label{sec:conclusion}

\chen{In this paper, we provide an in-depth analysis of the current state of security and privacy in FL systems, and we explore the core functionalities and architectures of these systems.}

\chen{First, we outline FL systems and their distinctions from traditional DML, and review the typical architecture layers of FL systems: user services, algorithms, and infrastructure. }

\chen{Next, we discuss the security and privacy challenges faced at each architectural layer of FL systems, identifying potential threats, including non-malicious failures and malicious attacks, along with a overview of possible attack methods. To address these threats, we analyze various existing defense mechanisms and emphasize the importance of building a robust risk management framework. This framework integrates evaluation, control, and monitoring phases to proactively identify and mitigate potential vulnerabilities within the system, thereby enhancing overall system resilience.}

\chen{Finally, we examine the diverse practical applications of FL, focusing on its use across multiple industries and the associated security and privacy challenges. These include applications in healthcare, finance, wireless communications, autonomous vehicles, recommendation systems, and smart city infrastructure. We also identify future research directions, such as the reliability and security of large models, Dynamic Adaptive Defense Techniques, system fault tolerance, and balancing privacy with model performance.}

\bibliographystyle{plain}
\bibliography{ref}

\end{document}